\documentclass{07akws-procs9x6}

\begin{document}

\title{Perspectives on Lorentz and CPT Violation}

\author{V.\ Alan Kosteleck\'y}

\address{Physics Department, Indiana University \\
Bloomington, IN 47405, U.S.A.}

\maketitle

\abstracts{
This talk offers some comments and perspectives 
on Lorentz and CPT violation.
}

\section{Introduction}

Lorentz symmetry is the invariance of physical laws
under rotations and boosts.
As a global symmetry over Minkowski spacetime,
it underlies the theory of Special Relativity 
and the Standard Model of particle physics,
where it is intimately tied to CPT invariance.
As a local symmetry of freely falling frames,
it is an essential component of General Relativity.
Nonetheless,
the possibility exists that nature may exhibit
tiny violations of Lorentz symmetry.
This talk presents some perspectives 
on the basic motivations and reasoning in this subject.

Since Lorentz symmetry has been verified in many experiments,
as has CPT invariance,
it is reasonable to ask why relativity violations 
are worth considering.
A sufficient theoretical motivation is the need for a 
consistent description of Lorentz and CPT violation 
to offer guidance for experimental tests.
However,
a stronger motivation 
is the prospect that Lorentz and CPT violation  
can serve as a sensitive potential signal
for physics at the Planck scale.
In fact,
the present interest in the subject was triggered 
by the realization that natural mechanisms 
for Lorentz and CPT violation
exist in unified theories at the Planck scale.\cite{ksp}
The large range of existing 
phenomenological and experimental activities
stems from the application of effective field theory\cite{kp95}
and the construction 
of the Standard-Model Extension (SME)\cite{ck,akgrav}
to catalogue and predict observable effects.

\section{Approaches and origins}

The study of Lorentz and CPT violation 
can be approached on three distinct levels.
First,
at the level of fundamental theory at the unification scale,
one can investigate possible mechanisms 
and determine their features and implications.
Second,
at the level of theory at accessible scales,
one can seek a description of the resulting effects
that is quantitative,
general,
and compatible with the established physics
of the Standard Model and General Relativity.
Finally,
at the level of observation and experiment,
one can study and perform 
both high sensitivity tests and broad searches,
preferably ones that are feasible
with existing or near-future technology.

The first point to establish is whether 
Lorentz violation can indeed occur in a fundamental theory.
Without at least one viable mechanism,
the interest of the idea would be much reduced. 
A plausible origin for Lorentz violation 
has been identified in string field theory,\cite{ksp} 
which has interactions with a generic structure 
that could in principle
trigger spontaneous violation of Lorentz symmetry
and generate vacuum expectation values
for Lorentz tensor fields.
More recently,
numerous other mechanisms for Lorentz violation
at the fundamental level have been proposed including,
for example,
ones involving 
noncommutative field theories,\cite{ncqed}
spacetime-varying fields,\cite{spacetimevarying}
quantum gravity,\cite{qg}
random-dynamics models,\cite{fn}
multiverses,\cite{bj}
brane-world scenarios,\cite{brane}
supersymmetry,\cite{susy}
and massive gravity.\cite{modgrav}

Given that mechanisms exist for Lorentz and CPT violation
in an underlying theory,
it is natural to ask about the consequences
for physics at observable scales. 
In particular,
the question arises as to the best method
to develop a description of the possible effects.

\section{Describing Lorentz violation}

Some key criteria offer valuable guidance 
in the search for a suitable theoretical framework
for describing Lorentz violation at attainable scales.
The first is {\it coordinate independence.}
It has been accepted since long before 1905 that 
the physics of a system should be independent 
of a change of observer coordinates.
This holds
whether a coordinate change is implemented
via a Lorentz transformation or in any other way. 

The second is {\it realism.}
Since 1905,
when virtually no fundamental particles were known
and quantum physics was at its dawn,
thousands of people have invested millions of person-hours
and billions of dollars
in establishing the Standard Model of particle physics
and General Relativity
as an accurate description of nature.
To be of real interest nowadays,
any proposed theoretical framework for Lorentz violation
must incorporate this well-established physics.

The third is {\it generality.}
No compelling evidence for Lorentz violation exists
at present.
Physics is therefore currently in the position 
of searching for a violation,
as opposed to attempting to understand an observed effect.
In the searching phase,
it is desirable to have the most general possible formulation
so that no region is left unexplored.
This is in strong contrast to the modeling phase,
where considerations such as simplicity are important
in attempts to understand a known effect.

Armed with these criteria,
we can follow the basic reasoning 
that leads to the application of effective field theory 
and the construction of the SME.

\subsection{Modified Lorentz transformations}

Since the essential content of Special Relativity 
is the idea that physics is invariant under
Lorentz transformations,
the most obvious approach to 
describing relativity violations
is to investigate modifications of the Lorentz transformations.
In fact,
the literature since 1905 abounds
with various {\it ad hoc} proposals of this type.
However,
independent of any specific proposal,
this approach has some serious disadvantages. 

One is that 
a textbook Lorentz transformation acts on the observer
and therefore corresponds merely to a change of coordinates,
i.e.,
a change of reference frame.
However,
according to the above criterion of coordinate independence,
a frame change cannot have physical implications by itself.
The key feature of Special Relativity
is really the requirement that the equations 
for the system being observed
must be covariant under a Lorentz transformation,
which intrinsically assumes that Lorentz symmetry is exact.
This approach is therefore problematic for investigating violations.

It is of course possible to construct special models
imposing form covariance of the system
under some {\it ad hoc} alternative transformation.
However,
any specific such proposal runs counter to the criterion
of generality.
Moreover,
some kinds of violations are difficult and perhaps 
even impossible to countenance via this approach.
For example,
Lorentz violation in nature might well be particle-species dependent,
but it is very challenging 
to formulate a description of this flavor dependence 
based on modified Lorentz transformations of the observer.
The criterion of realism presents a further substantial obstacle,
since it is awkward at best to implement such models 
in the context of the Standard Model and General Relativity.

\subsection{Modified dispersion laws}

The above discussion suggests that
a general and realistic investigation of Lorentz violation
is most naturally performed directly
in terms of the properties of a system 
rather than via modifications of the Lorentz transformations.
A simple implementation of this 
is to study modifications of particle dispersion laws.
However,
this also suffers from serious drawbacks.

One issue involves the criterion of generality.
Modifications of dispersion laws
can only describe changes in the free propagation of particles
and perhaps also partially account for interaction kinematics.
However,
physics is far more than free propagation,
and this approach therefore disregards 
a large range of interesting Lorentz-violating effects
involving interactions.

There are also various issues associated 
with the choice of modifications to the dispersion law.
Not all choices are compatible 
with desirable features such as originating from an action.
Also,
meaningful physical measurements must necessarily compare
two quantities,
so some choices may be unphysical.
In particular,
calculations with a modified dispersion law
that yield apparent changes of properties
are insufficient by themselves
to demonstrate physical Lorentz violation.
A simple example of a modified dispersion law 
with no observable consequences in Minkowski spacetime
is\cite{ck}
$p^\mu p_\mu = m^2 + a_\mu p^\mu$,
where $a_\mu$ is a prescribed set of four numbers 
in a given frame. 
Direct calculations with this dispersion law appear to give
Lorentz-violating properties that depend 
on the preferred vector $a_\mu$,
but in fact they are unobservable 
because $a_\mu$ can be eliminated 
via a physically irrelevant redefinition 
of the energy and momentum. 
The observability of modifications to a dispersion law
can be challenging to demonstrate.

\subsection{Effective field theory and the SME}

We see that the desiderata
for a satisfactory description of Lorentz violation
include a comprehensive treatment of free and interacting effects 
in all particle species.
Remarkably,
a model-independent and general approach 
of this type exists.

The key is to take advantage of the idea 
that Lorentz violation at attainable energies
can be described using effective field theory,
independent of the underlying mechanism.\cite{kp95,kleh}
Starting from the Standard Model coupled to General Relativity,
we can add to the action all possible scalar terms 
formed by contracting operators for Lorentz violation
with coefficients that control the size of the effects.
The operators are naturally ordered according to their mass dimension.
The resulting realistic effective field theory
is the SME.\cite{ck,akgrav}
Since CPT violation in realistic field theories
comes with Lorentz violation,\cite{owg}
the SME also incorporates general CPT violation.

By virtue of its construction,
the SME satisfies the three guiding criteria
of coordinate invariance, realism, and generality.
Moreover,
it handles simultaneously all particle species,
including both propagation and interaction properties,
so its equations of motion contain
all action-compatible modifications of realistic dispersion laws. 
The coordinate invariance implies 
that physics is unaffected by observer frame changes,
including Lorentz and other transformations,
while particle transformations can produce observable effects
of Lorentz violation.

The primary disadvantage of the SME approach 
is its comparative complexity 
and the investment required to become proficient with its use. 
However,
this is outweighed by its advantages
as a realistic, general, and calculable framework
for describing Lorentz violation.
Judicious choices establishing relations
among the SME coefficients for Lorentz violation 
yield elegant and simple models 
that can serve as a theorist's playground,
while the general case offers guidance 
for broad-based experimental searches.

\section{Gravity and Lorentz violation}

The SME allows for both global\cite{ck} 
and local\cite{akgrav} Lorentz violation,
and interesting effects arise from
local Lorentz violation in the gravitational context.
In general,
local Lorentz violation 
can be understood as arising
when a nonzero coefficient $t_{abc\ldots}$ 
for Lorentz violation
exists in local freely falling frames.\cite{akgrav}
The cofficient $t_{abc\ldots}$
can be converted to a coefficient
$t_{\lambda\mu\nu\ldots}$ on the spacetime manifold
using the vierbein $e_\mu^{\phantom{\mu}a}$.

One result is that spontaneous violation of local Lorentz symmetry 
is always accompanied by spontaneous diffeomorphism violation,
and vice versa.\cite{bk}
A nonzero coefficient $t_{abc\ldots}$ 
is the vacuum value of a local Lorentz tensor field,
and it implies Lorentz violation 
because it is invariant 
instead of transforming like a tensor
under particle transformations.
The vierbein ensures that 
there is a corresponding spacetime tensor field
with vacuum value $t_{\lambda\mu\nu\ldots}$
on the spacetime manifold,
which in turn implies spontaneous diffeomorphism breaking 
because it is invariant 
instead of transforming like a tensor
under particle diffeomorphisms.

A more surprising result is that 
explicit Lorentz violation is generically incompatible
with Riemann geometry.\cite{akgrav}
Explicit violation occurs when the
SME coefficients are externally prescribed,
but the ensuing equations of motion
turn out to be inconsistent with the Bianchi identities.
This result also holds in Riemann-Cartan spacetime.
However,
spontaneous violation evades the difficulty
because it generates the SME coefficients dynamically, 
thereby ensuring compatibility with the underlying spacetime geometry.

Spontaneous local Lorentz violation 
is accompanied by up to 10 Nambu-Goldstone (NG)
modes.\cite{bk}
With a suitable choice of gauge,
these modes can be identified with components of the vierbein
normally associated
with local Lorentz and diffeomorphism gauge freedoms.
The physical role of the NG modes varies,
but in general they represent long-range forces
that can be problematic for phenomenology.
However,
in certain models the NG modes can be interpreted as photons,
thus offering the intriguing prospect 
that the existence of light could be a consequence
of Lorentz violation
instead of local U(1) gauge invariance,
with concomitant observable signals.\cite{bk,baak}
A similar interpretation is possible for the 
graviton.\cite{kpgr}
Other potential experimental signals 
arise from NG modes in the gravity\cite{bak}
and matter\cite{kt} sectors,
and from torsion.\cite{krtorsion}
The spectrum of vacuum excitations 
typically also includes massive modes
that may lead to additional observable effects.\cite{bfks}

\section{The search for signals}

The SME predicts some unique signals,
such as rotational, sidereal, and annual variations.
The effects are likely to be heavily suppressed,
perhaps as some power of the ratio 
of an accessible scale to the underlying scale,
but they could be detected 
using sensitive tools such as interferometry.
For example,
meson interferometry offers the potential to identify 
flavor- and direction-dependent energy shifts
of mesons relative to antimesons,\cite{mesons}
while exquisite interferometric sensitivity  
to polarization-dependent effects of photons
is attained using cosmological birefringence.\cite{photon}
Conceivably, 
SME effects might even be reflected in existing data,
such as those for  
flavor oscillations of neutrinos.\cite{neutrinos}
Overall,
an impressive range of sensitivities
in the matter, gauge, and gravitational sectors of the SME 
has been achieved.\cite{kr}

Despite a decade of intense activity,
most of the SME coefficient space 
is still unexplored by experiments,
and many basic theoretical issues are unaddressed.
The study of relativity violations remains fascinating, 
with the enticing prospect 
of identifying a signal from the Planck scale.

\section*{Acknowledgments}

This work was supported in part
by DoE grant DE-FG02-91ER40661 and
NASA grant NAG3-2914.


\begin{thebibliography}{xx}
\def\etal {{\it et al.}}

\bibitem{ksp}
V.A.\ Kosteleck\'y and S.\ Samuel,
Phys.\ Rev.\ D {\bf 39}, 683 (1989);
V.A.\ Kosteleck\'y and R.\ Potting,
Nucl.\ Phys.\ B {\bf 359}, 545 (1991).

\bibitem{kp95}
V.A.\ Kosteleck\'y and R.\ Potting,
Phys.\ Rev.\ D {\bf 51}, 3923 (1995).

\bibitem{ck}
D.\ Colladay and V.A.\ Kosteleck\'y,
Phys.\ Rev.\ D {\bf 55}, 6760 (1997);
Phys.\ Rev.\ D {\bf 58}, 116002 (1998).

\bibitem{akgrav}
V.A.\ Kosteleck\'y,
Phys.\ Rev.\ D {\bf 69}, 105009 (2004).

\bibitem{ncqed}
See, for example,
I.\ Mocioiu \etal,
Phys.\ Lett.\ B {\bf 489}, 390 (2000);
S.M.\ Carroll \etal,
Phys.\ Rev.\ Lett.\ {\bf 87}, 141601 (2001).

\bibitem{spacetimevarying}
V.A.\ Kosteleck\'y \etal,
Phys.\ Rev.\ D {\bf 68}, 123511 (2003).

\bibitem{qg}
See, for example,
G.\ Amelino-Camelia \etal, 
AIP Conf.\ Proc.\ {\bf 758}, 30 (2005);
N.E.\ Mavromatos,
Lect.\ Notes Phys.\ {\bf 669}, 245 (2005);
Y.\ Bonder and D.\ Sudarsky,
arXiv:0709.0551.

\bibitem{fn}
C.D.\ Froggatt and H.B.\ Nielsen,
hep-ph/0211106.

\bibitem{bj}
J.D.\ Bjorken,
Phys.\ Rev.\ D {\bf 67}, 043508 (2003).

\bibitem{brane}
See, for example,
C.P.\ Burgess \etal,
JHEP {\bf 0203}, 043 (2002).

\bibitem{susy}
M.\ Berger and V.A.\ Kosteleck\'y,
Phys.\ Rev.\ D {\bf 65}, 091701(R) (2002);
P.A.\ Bolokhov \etal,
Phys.\ Rev.\ D {\bf 72}, 015013 (2005).

\bibitem{modgrav}
See, for example,
G.\ Dvali \etal, 
Phys.\ Rev.\ D {\bf 76}, 044028 (2007);
D.S.\ Gorbunov and S.M.\ Sibiryakov,
JHEP {\bf 0509}, 082 (2005);
M.V.\ Libanov and V.A.\ Rubakov,
JHEP {\bf 0508}, 001 (2005);
N.\ Arkani-Hamed \etal, 
JHEP {\bf 0507}, 029 (2005);
V.A.\ Kosteleck\'y and S.\ Samuel,
Phys.\ Rev.\ D {\bf 42}, 1289 (1990);
Phys.\ Rev.\ Lett.\ {\bf 66}, 1811 (1991).

\bibitem{kleh}
V.A.\ Kosteleck\'y and R.\ Lehnert,
Phys.\ Rev.\ D {\bf 63}, 065008 (2001).

\bibitem{owg}
O.W.\ Greenberg,
Phys.\ Rev.\ Lett.\ {\bf 89}, 231602 (2002).

\bibitem{bk}
R.\ Bluhm and V.A.\ Kosteleck\'y,
Phys.\ Rev.\ D {\bf 71}, 065008 (2005).

\bibitem{baak}
B.\ Altschul and V.A.\ Kosteleck\'y,
Phys.\ Lett.\ B {\bf 628}, 106 (2005).

\bibitem{kpgr}
V.A.\ Kosteleck\'y and R.\ Potting,
Gen.\ Rel.\ Grav.\ {\bf 37}, 1675 (2005).

\bibitem{bak}
Q.G.\ Bailey and V.A.\ Kosteleck\'y,
Phys.\ Rev.\ D {\bf 74}, 045001 (2006).

\bibitem{kt}
V.A.\ Kosteleck\'y and J.D.\ Tasson,
in preparation.

\bibitem{krtorsion}
V.A.\ Kosteleck\'y, N.\ Russell, and J.D.\ Tasson,
arXiv:0712.4393.

\bibitem{bfks}
R.\ Bluhm \etal,
arxiv:0712.4119;
V.A.\ Kosteleck\'y and S.\ Samuel,
Phys.\ Rev.\ D {\bf 40}, 1886 (1989);
Phys.\ Rev.\ Lett.\ {\bf 63}, 224 (1989).

\bibitem{mesons}
H.\ Nguyen (KTeV),
hep-ex/0112046;
A.\ Di Domenico \etal\ (KLOE),
these proceedings;
B.\ Aubert \etal\ (BaBar),
hep-ex/0607103;
arXiv:0711.2713;
D.P.\ Stoker (BaBar), 
these proceedings;
J.M.\ Link \etal\ (FOCUS),
Phys.\ Lett.\ B {\bf 556}, 7 (2003);
V.A.\ Kosteleck\'y,
Phys.\ Rev.\ Lett.\ {\bf 80}, 1818 (1998);
Phys.\ Rev.\ D {\bf 61}, 016002 (2000);
Phys.\ Rev.\ D {\bf 64}, 076001 (2001).

\bibitem{photon}
V.A.\ Kosteleck\'y and M.\ Mewes,
Phys.\ Rev.\ Lett.\ {\bf 87}, 251304 (2001);
Phys.\ Rev.\ D {\bf 66}, 056005 (2002);
Phys.\ Rev.\ Lett.\ {\bf 97}, 140401 (2006);
Phys.\ Rev.\ Lett.\ {\bf 99}, 011601 (2007).

\bibitem{neutrinos}
L.B.\ Auerbach \etal,
Phys.\ Rev.\ D {\bf 72}, 076004 (2005);
B.J.\ Rebel and S.F.\ Mufson,
these proceedings;
V.A.\ Kosteleck\'y and M.\ Mewes,
Phys.\ Rev.\ D {\bf 69}, 016005 (2004);
Phys.\ Rev.\ D {\bf 70}, 031902(R) (2004);
Phys.\ Rev.\ D {\bf 70}, 076002 (2004);
T.\ Katori \etal, 
Phys.\ Rev.\ D {\bf 74}, 105009 (2006);
V.\ Barger \etal,
Phys.\ Lett.\ B {\bf 653}, 267 (2007);
K.\ Whisnant, these proceedings.

\bibitem{kr}
Results are tabulated in 
V.A.\ Kosteleck\'y and N.\ Russell,
arXiv:0801.0287.

\end{thebibliography}
\end{document}